# Observation of the Crossover from Two-Gap to Single-Gap Superconductivity through Specific Heat Measurements in Neutron Irradiated MgB$_2$

M.Putti[1], M.Affronte[2], C.Ferdeghini[1], C.Tarantini[1], E.Lehmann[3]

[1]CNR-INFM-LAMIA and Dipartimento di Fisica, Università di Genova, Via Dodecaneso 33, 16146 Genova, Italy

[2]CNR-INFM-S3 and Dipartimento di Fisica, Università di Modena e Reggio Emilia
Via G.Campi 213/A, I-41100 Modena, Italy

[3]Paul Scherrer Institut, CH-5232 Villigen, Switzerland

**Abstract**

We report specific heat measurements in neutron irradiated MgB$_2$ samples, for which the critical temperature has been suppressed down to 8.5 K, but the superconducting transition remains extremely sharp, indicative of a defect structure extremely homogeneous. Our results demonstrate that the two-gap feature is evident in the temperature range above 21 K, while the single-gap superconductivity is well established as a bulk property not associated to local disorder fluctuations when T$_c$ is decreased down to 11 K.

The two-gap nature of superconductivity in $MgB_2$ is a unique feature of this material that stimulates many theoretical and experimental investigations. *Ab initio* calculations[1-3] have showed that $MgB_2$ has two weakly coupled gaps, $\Delta_\sigma(0) \approx 7$ meV and $\Delta_\pi(0) \approx 2$ meV, residing on disconnected sheets of the Fermi surface formed by in-plane $p_{xy}$ boron orbitals (the σ-bands) and out-of-plane $p_z$ boron orbitals (the π-bands). The two-gap Eliashberg theory accounts for many anomalies in superconducting and normal properties of pure $MgB_2$ , but still some aspects need to be clarified. Golubov and Mazin[4] pointed out that the two-gap superconductivity must be strongly affected by disorder. In particular, inter-band scattering by non-magnetic impurities is expected to induce pair breaking which suppresses the critical temperature $T_c$ down to about 20 K, where an equivalent one-gap BCS system with isotropic coupling stabilizes. The pair breaking mechanism is expected to induce the increase of the small gap and the decrease of the large one, so these are to merge to the BCS value once a complete isotropization has taken place. Whilst the suppression of superconductivity may be driven by different effects, for instance charge doping and the consequent filling of electronic σ bands,[5] the merging of the two gaps is a peculiar result of the inter-band scattering. So, a direct observation of the crossover from two- to single-gap superconductivity will provide the final "smoking gun" evidence for the two-band model.[6]

Several efforts have been done to evaluate the energy gaps in samples where the defects were introduced by substitution (Al in sites of Mg,[7-10] C in sites of B[11-16]), and by neutron irradiation,[17] and in disordered thin films.[18] Despite the variety of techniques used to induce disorder and the different experimental technique employed to estimate the gaps (tunneling, point contact spectroscopy, specific heat, photoemission spectroscopy), it turns out that the two gaps $\Delta_\pi$ and $\Delta_\sigma$ scale in a quite general way with the critical temperature of the sample. The gap associated with the π-band is only weakly sensitive to the critical temperature, while $\Delta_\sigma$ decreases almost linearly, so that the two gaps tend to join together. This behaviour is rather well established for $T_c$ values ranging from 39 K to 25-30 K, while, when $T_c$ approaches 20 K, the results are less clear. In Al-doped samples there is no evidence of single gap superconductivity.[7-10] In C-doped samples several experiments[11-15] suggest that the extrapolated energy gaps do not merge, except for one C-doped single crystal with $T_c$=19 K for which single-gap superconductivity was claimed.[16] Data on undoped and disordered $MgB_2$ are not available below 20 K, the temperature range where a single gap is expected.

The lack of unambiguous evidences of the merging of the gaps is disappointing, but some explanations can be considered. In Al- and C-doped samples the suppression of $T_c$ is due to at least two concomitant mechanisms, i.e. σ band filling and inter-band scattering; when $T_c$ is decreased to 20 K the inter-band scattering might be not strong enough to induce the merging of the gaps.[5] Moreover, heavily doped compounds become structurally unstable and present inhomogeneous properties, inadequate to check such a critical point.

To overcome these problems we propose the study of samples in which disorder has been introduced by neutron irradiation, which does not induce remarkable changes in the band structure. In this letter we report direct observation of the crossover from two- to single-gap superconductivity in neutron irradiated $MgB_2$ samples where the critical temperature was suppressed down to 8.5 K. The irradiation procedure we used produces samples with defect structure extremely homogeneous so that the superconducting transition remains extremely sharp even in the heavily irradiated samples. Our results demonstrate that the two-gap feature is evident in the temperature range above 21 K, while the single-gap superconductivity is well established as a bulk property not associated to local disorder fluctuations when $T_c$ is suppressed down to 11 K.

Polycrystalline $MgB_2$ samples, prepared by direct synthesis from Mg and isotopically enriched $^{11}B$, were irradiated at the spallation neutron source SINQ (thermal neutron flux density up to $1.6 \cdot 10^{13}$ $cm^{-2} s^{-1}$) at the Paul Sherrer Institut (PSI). The thermal neutron fluence was varied from $10^{17}$ to $1.4 \times 10^{20}$ $cm^{-2}$. In ref.19 we demonstrated that the most important damage mechanism in our samples consists in neutron capture reactions by the residual $^{10}B$ present (less than 0.5%). Due to the large cross section of this reaction, the neutron penetration in bulk samples rich in $^{10}B$ is partially shielded (natural boron consists of 80% of $^{11}B$ and 20% of $^{10}B$).[20] Thus, to avoid self-shielding effects, we irradiated $Mg^{11}B_2$ samples for which the penetration depth of thermal neutrons is much deeper than the sample thickness so that nuclear reactions take place homogeneously inside the sample. The defects in the form of atom displacements are created by the recoil of $^4He$ and $^7Li$, produced by the nuclear reaction, emitted in isotropic directions. This makes the defect distribution very homogeneous.

Resistivity value at 40 K, $\rho(40)$, residual resistivity ratio, RRR, crystallographic axes, $T_c$ and width of superconductig transition, $\Delta T_c$, of a sample series are reported in table I. $\rho(40)$, RRR, crystallographic axes and $T_c$ scale monotonously with the thermal neutron dose. In particular, as the fluence increases, $\rho(40)$ increases by two order of magnitude, the a- and c-axes increase by 0.4% and 1%, respectively, and $T_c$ decreases down to 8.5 K. On the other hand, $\Delta T_c$ remains very narrow in all the irradiated samples.

Fig. 1 shows the critical temperature as a function of $\rho(40)$. $T_c$ linearly decreases with $\rho(40)$ in the whole range, in agreement with what observed in $^4He$ irradiated $MgB_2$ thin films.[21] Notice that, even though the few data available around 20 K are not sufficient to infer any peculiar behaviour, it is clear that our data, as well as those of Ref. 21, do not show any $T_c$ saturation. This suggests that pair breaking due to inter-band scattering is not the only mechanism which suppresses superconductivity in irradiated MgB$_2$ samples. On the other hand, the roughly linear scaling of T$_c$ with ρ(40) indicates that a strong correlation exists between the mechanism increasing ρ(40) (scattering with atomic scale defects) and that suppressing superconductivity. This behaviour is common to other superconductors. In amorphous transition metals and damaged A15 superconductors, for instance, a smearing of the peak in the electron density of states at the Fermi level was suggested;[22] yet this mechanism does not simply apply to MgB$_2$, whose density of states is rather flat around the Fermi level, so that other mechanisms which suppress the electron-phonon coupling should be invoked.

The energy gaps have been estimated by analysing specific heat $c$ data. Such measurements probe the bulk properties of materials and are not affected by local fluctuations of disorder.

Specific heat measurements were performed on the six samples listed in table II by a Quantum Design PPMS-7T which makes use of the relaxation method. For each sample we performed a set of measurements between 2 and 40 K and in magnetic field of 0 and 7 T. Addenda with the exact amount of Apiezon N grease used for thermal contact were carefully measured before each run.

The T$_c$ and ΔT$_c$ values estimated from specific heat are reported in table II; they are quite close to the values estimated from susceptibility (see table I) . In fig.2 we plot the zero field data as $c/T$-vs-$T^2$ and, for sake of clarity, only the results for P0, P3.5, P3.7 and P6 samples are shown. Note that the superconducting transition is remarkably sharp in all the irradiated samples.[23] It is worth comparing the curves in fig. 2 with the analogous ones obtained in Al-doped MgB$_2$.[7-10] In Al-doped samples the superconducting transition becomes broader and broader as the doping increases, while for the irradiated samples the transitions remain sharp, whatever the level of disorder introduced.

In order to extract information on the superconducting energy gaps, the reduced electronic specific heat, $c_{sc}/\gamma T$, has been calculated (here γ is the Sommerfeld constant) by following the usual procedure described in ref. 7. The results for the samples P0, P4, P5, P6 are reported in fig. 3 as a function of the reduced temperature $t$=T/T$_c$ and γ values are reported in Table II for all the samples.

The specific heat in the superconducting state is analyzed within the two-band α-model[24] that was widely used to analyze the experimental specific heat data of pure and doped MgB$_2$. Recently, Dolgov et al.[25] showed that, in spite of the strong modifications of the density of states by interband scattering, the α-model is sufficiently accurate to extract gap values from specific heat of disordered samples. Within this model, the σ and π bands contribute to the specific heat proportionally to the γ$_σ$/ γ and γ$_π$/ γ fractions, respectively, γ$_σ$ and γ$_π$ being the Sommerfeld coefficient of the σ and π bands. The evidence of the π–gap comes from the excess of $c_{sc}/\gamma T$ at $t$~0.2 which cannot be accounted for by the σ band contribution alone since, in this $t$-range, it falls down exponentially.

Such excess is well evident in the pristine sample (P0), but disappears in the most irradiated ones (P5 and P6). This guess is substantiated by data fitting analysis. Therefore, we used two fitting procedures: within a two-gap framework, three free parameters, $\alpha_\sigma=\Delta_\sigma(0)/k_BT_c$, $\alpha_\pi=\Delta_\pi(0)/k_BT_c$, and $x=\gamma_\pi/\gamma$, were introduced whilst within a single-gap framework, only one free parameter, $\alpha=\Delta(0)/k_BT_c$, was used. For the samples P0, P3.5, P3.7, P4, the experimental data can only be reproduced by considering two gaps, while the single-gap analysis reproduces quite well the electronic specific heat of P5 and P6. In fig. 3 we report the best fitting curves as continuous lines. The sample P4, which presents $T_c=21$ K, close to the value predicted for the merging of the gaps, was carefully investigated and the best fit curve obtained by a single-gap analysis is reported in fig. 3 for comparison (dotted line). It is clear that the single-gap curve cannot reproduce the feature of the experimental data, and we conclude that the for this sample the merging of the gaps has not occurred yet. The best fit parameters for all the samples are reported in table II.

In fig.4, $\Delta_\sigma(0)$, $\Delta_\pi(0)$ and $\Delta(0)$ are plotted as a function of $T_c$. We may easily distinguish two regions: for $T_c \geq 21$ K, the two gap-feature is observed, while for samples with $T_c<20$K, superconductivity is characterized by a single-gap. In the two-gap region we find that $\Delta_\sigma(0)$ decreases almost linearly with $T_c$, while $\Delta_\pi(0)$ remains nearly constant. A close inspection shows that, out of the error bars, as $T_c$ decreases, $\Delta_\pi(0)$ slightly rises, showing a flat maximum around 30 K, and then decreases; this trend is confirmed by $\Delta_\pi(0)$ values obtained on neutron irradiated MgB$_2$ by Wang et al.,[17] showed in the same figure for comparison. This behaviour could be associated with the compensation of the inter-band scattering, which increases $\Delta_\pi(0)$, and with the diminution of electron-phonon coupling driven by the disorder, as discussed in ref. 5.

In the single-gap region, not investigated before, $\Delta(0)$ decreases with $T_c$ reaching a value of 1 meV at $T_c=8.7$ K. Our results fix the $T_c$ value at which the merging of the gaps occurs between 11 and 21 K. Since the gaps are still well separated at 21 K, the merging has to take place at a temperature lower than the 20 K predicted for isotropic MgB$_2$. Recently Wilke et al.,[26] analysing the upper critical field of neutron irradiated samples, suggested that the bands become fully mixed only when $T_c$ is near 10 K. These results do not contradict the two-band model. As discussed before, the lack of $T_c$ saturation at 20 K suggests that disorder not only increases the inter-band scattering, but also affects the electron-phonon coupling. On the other hand, the observation of single-gap superconductivity unambiguously indicates that inter-band scattering has effectively lead to the merging of the gaps.

Interestingly, in the single-gap regime, the reduced gap values, $2\Delta(0)/k_BT_c$, are 2.8-2.7, lower than the BCS value 3.52. Reduced gap values lower than BCS one were observed also in conventional disordered superconductor (alloyed and irradiated samples).[27] If we are allowed to exclude inhomogeneity effects, breakdowns of the BCS theory in the case of highly disordered systems should be considered. This is an intriguing point, not approached systematically in the past, which will require further investigations.

In conclusion, we succeeded in introducing defects in pure $Mg^{11}B_2$ by thermal neutron irradiation. The critical temperature has been suppressed down to 8.5 K, but the superconducting transitions remain sharp. The electronic specific heat of these samples has been analyzed within the α-model in order to study the crossover from two-gap to single-gap superconductivity. For $T_c$ values above 21 K the two-gap feature remains evident, while we have proved that below 11 K the single-gap superconductivity is completely established. So we fix between 11 and 21 K the $T_c$ values at which the merging of the gaps occurs, in agreement with the prediction of the two-band theory. What remains to be understood is the mechanism that suppresses the superconductivity in heavily disordered $MgB_2$. In fact, the inter-band scattering alone cannot account for such low $T_c$ values, and the suppression of electron-phonon coupling should be carefully investigated.

**Figure captions**

Figure 1. $T_c$ as a function of $\rho(40)$ for the samples listed in table I. The continuous line is a guide for the eye.

Figure 2. c/T as a function of $T^2$ for the samples P0, P35, P37 and P6.

Figure 3. $c_{sc}/\gamma T$ as a function of $t$ for the samples P0, P4, P5 and P6. The best fit curves are plotted as continuous lines. For the sample P4 the best fit curves obtained within the single-gap (dotted line) and two-gap (continuous line) analyses are presented for comparison.

Figure 4. $\Delta_\sigma(0)$ (filled symbols), $\Delta_\pi(0)$ (empty symbols) and $\Delta(0)$ (half-filled symbols) as a function of $T_c$. Triangle symbols represent data of ref. [17].

**Table captions**

Table I. Thermal neutron fluence, $\Phi$, resistivity value at 40 K, $\rho(40)$, residual resistivity ratio, RRR=$\rho(300)/\rho(40)$, crystallographic axes, $a$ and $c$, $T_c$ and width of superconductig transition, $\Delta T_c$, of a sample series. The critical temperature $T_c$ was determined by susceptibility measurements and $\Delta T_c$ was estimated as the difference between $T_c$ estimated at the 10% and 90% of the transition.

Table II. $T_c$ and $\Delta T_c$ estimated by specific heat; the Sommerfeld coefficient $\gamma$, estimated by fitting the normal state experimental data measured at 7 T as c/T vs $T^2$; the best fit parameters $2\alpha_\sigma=2\Delta_\sigma(0)/k_B T_c$, $2\alpha_\pi=2\Delta_\pi(0)/k_B T_c$, and $x=\gamma_\pi/\gamma$ and the gap values $\Delta_\sigma(0)$ and $\Delta_\pi(0)$ evaluated for the samples P0, P3.5, P3.7, P4; the best fit parameter $2\alpha=2\Delta(0)/k_B T_c$ and the gap value $\Delta(0)$ evaluated for the samples P5, P6.

Table I

| Samples | Φ (cm$^{-2}$) | ρ(40) (μΩcm) | RRR | a (Å) | c (Å) | $T_c$ (K) | $\Delta T_c$ (K) |
|---|---|---|---|---|---|---|---|
| **P0** | 0 | 1.6 | 11.1 | 3.084 | 3.519 | 38.8 | 0.3 |
| **P1** | 1.0·10$^{17}$ | 2.4 | 6.9 | 3.083 | 3.524 | 38.35 | 0.3 |
| **P2** | 6.0·10$^{17}$ | 6.5 | 3.0 | 3.085 | 3.529 | 37.3 | 0.3 |
| **P3** | 7.6·10$^{17}$ | 16 | 2.0 | 3.083 | 3.525 | 35.6 | 0.3 |
| **P3.5** | 2.0·10$^{18}$ | 26 | 1.6 | 3.088 | 3.537 | 32.6 | 0.5 |
| **P3.7** | 5.5·10$^{18}$ | 41 | 1.2 | 3.088 | 3.538 | 25.8 | 0.6 |
| **P4** | 1.0·10$^{19}$ | 64 | 1.2 | 3.088 | 3.549 | 20.7 | 0.9 |
| **P5** | 3.9·10$^{19}$ | 124 | 1.1 | 3.095 | 3.558 | 11.0 | 0.4 |
| **P6** | 1.4·10$^{20}$ | 130 | 1.1 | 3.093 | 3.558 | 8.5 | 0.5 |

Table II

| Samples | $T_c$ (K) | $\Delta T_c$ (K) | γ (mJ/mol K$^2$) | $\gamma_\sigma/\gamma$ | $2\Delta_\sigma(0)/k_BT_c$ | $2\Delta_\pi(0)/k_BT_c$ | $\Delta_\sigma(0)$ meV | $\Delta_\pi(0)$ meV |
|---|---|---|---|---|---|---|---|---|
| **P0** | 38.5 | 0.8 | 3.0±0.1 | 0.54±0.05 | 3.8±0.1 | 1.07±0.06 | 6.3±0.3 | 1.8±0.1 |
| **P3.5** | 33.0 | 1.0 | 3.0±0.1 | 0.47±0.03 | 3.7±0.5 | 1.6±0.2 | 5.3±0.2 | 2.3±0.1 |
| **P3.7** | 26.0 | 1.0 | 3.0±0.1 | 0.56±0.05 | 3.5±0.3 | 1.70±0.1 | 3.9±0.4 | 1.9±0.2 |
| **P4** | 21.0 | 0.8 | 2.4±0.2 | 0.58±0.08 | 3.5±0.2 | 1.7±0.1 | 3.1±0.3 | 1.6±0.2 |
| **P5** | 11.0 | 0.6 | 2.5±0.2 | | 2.7±0.1 | | 1.3±0.1 | |
| **P6** | 8.7 | 0.4 | 2.4±0.2 | | 2.8±0.1 | | 1.1±0.1 | |

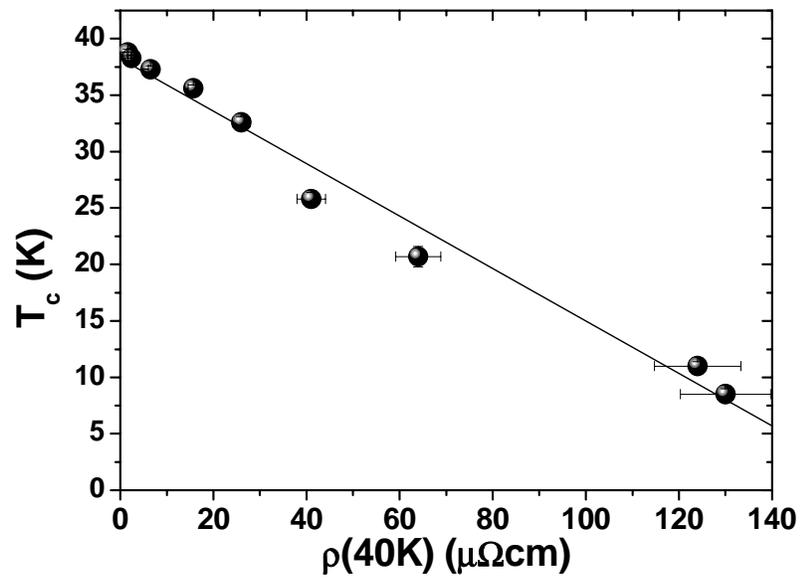

Figure 1

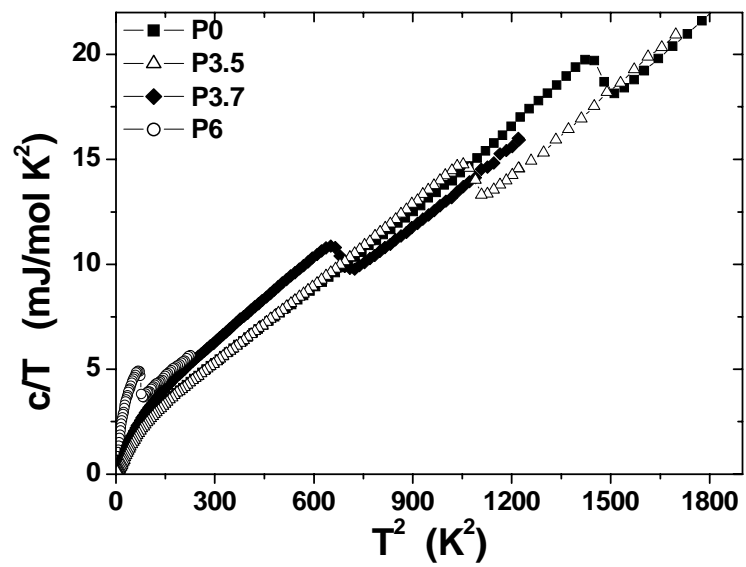

Figure 2

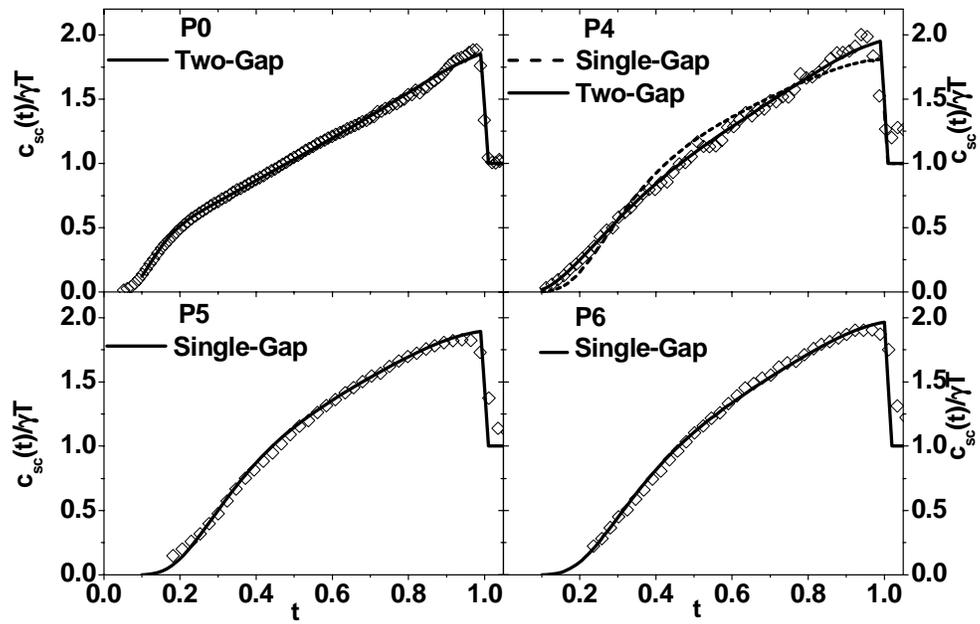

Figure 3

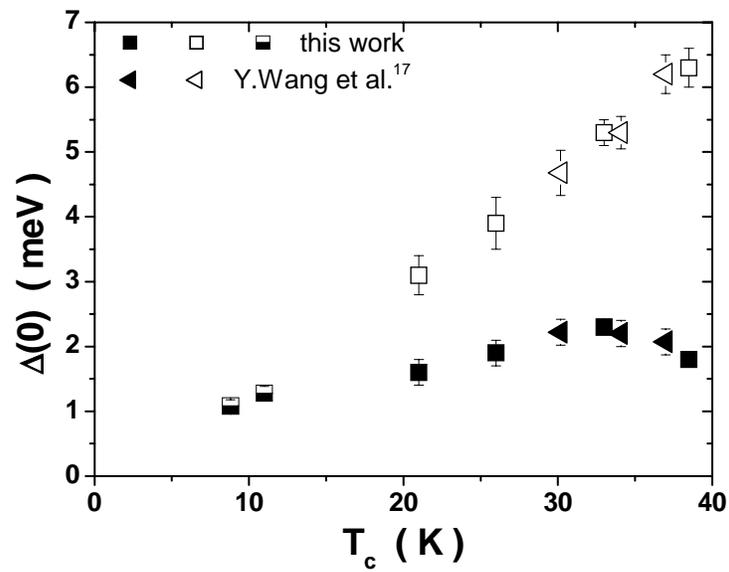

Figure 4